\documentstyle[aps,eqsecnum,multicol,epsf]{revtex}
\begin{document}
\draft

\title{Non-simultaneity in two-photon coincidence spectroscopy}
\author{L.\ Horvath, 
	B.\ C.\ Sanders	and B.\ F.\ Wielinga}
\address{
	Department of Physics, Macquarie University,
	Sydney, New South Wales 2109, Australia
}
\date{\today}
\maketitle

\begin{abstract}
Photon coincidence spectroscopy relies on detecting multiphoton
emissions from the combined atom-cavity system in atomic
beam cavity quantum electrodynamics experiments.
These multiphoton emissions from the cavity are nearly simultaneous
approximately on the cavity lifetime scale.
We determine the optimal time for the detection window of photon pairs
in two-photon coincidence spectroscopy.
If the window time is too short, some photon pairs will not be detected;
if the window time is too long, too many nearly coincident 
independent single photons will be falsely interpreted as being a photon pair.
\end{abstract}


\section{Introduction}
\label{intro}
Cavity quantum electrodynamics (CQED) has continued to develop rapidly, 
driven both by recent experimental successes and by the promise 
of exciting new applications. Advances in atom cooling techniques, as well
as development of high-Q optical cavities with large dipole coupling, 
have enabled testing of the strong-coupling regime of 
CQED~\cite{Thompson98}. Single-atom experiments are 
now feasible~\cite{Hood98,Mabuchi98}, and the possibility of trapping
atoms in optical cavities is tantalisingly close~\cite{Parkins}. 
Potential applications include quantum logic gates \cite{Turchette95}.

Applications of CQED rely critically on the quantum effects, namely the
entanglement between the field degree of freedom and the internal
electronic state of the atom~\cite{Carmichael96,Brune96}. 
This entanglement is not only challenging to achieve, it is also 
difficult to probe. In the optical regime of CQED, photon coincidence
spectroscopy (PCS) has been proposed as a feasible and unambiguous method
for detecting genuine quantum effects in CQED. This technique
employs a multichromatic driving field acting on the
combined atom-cavity system and detects multiphoton decays by detecting
photon coincidences in the cavity output 
field~\cite{Carmichael96,Sanders97,Horvath99}.

A difficulty arises in determining whether emitted photons are
coincident or not. Let us consider a single two-level atom (2LA) coupled to
a single mode of an optical cavity, and~$\omega$ is the angular frequency of
both the cavity mode and the 2LA. Multi-atom effects can be ignored provided
that the atomic density is sufficiently small~\cite{Carmichael99}.
In the electric dipole and
rotating-wave approximations, the Hamiltonian is 
\begin{equation} 
\label{JCH} 
H(g) = \omega (\sigma_z + a^{\dagger} a) 
	+ i g ( {\bf r} ) (a^{\dagger} \sigma_- - a \sigma_+ ) 
\end{equation} 
with~$\bf r$ the position of the atom,
$g({\bf r})$ the position-dependent dipole coupling strength, 
$a$ and $a^{\dagger}$ the annihilation and creation operators 
for photons in the cavity field, 
$\sigma_+$, $\sigma_-$, and $\sigma_z$ the 2LA  
raising, lowering and inversion operators, respectively, 
and $\hbar = 1$. 
The spectrum for this Hamiltonian is depicted in Fig.\ \ref{fig:ladder}
and is the well-known Jaynes-Cummings spectrum, or `ladder'~\cite{Jaynes63}.
The `dressed states' of the combined atom-cavity system
are designated by the lowest-energy state
\begin{equation}
\label{dressed:ground}
|0\rangle \equiv |0 \rangle_{\rm cav} \otimes |{\tt g} \rangle_{\rm atom} ,
\end{equation}
and, for $n$ a positive integer,
\begin{equation}
\label{dressed:excited}
|n \rangle_{\pm} \equiv \frac{\imath}{\sqrt{2}}
\left( |n-1\rangle_{\rm cav} \otimes |{\tt e}_{\rm atom} \rangle
	\pm \imath |n\rangle_{\rm cav}\otimes
	|{\tt g}\rangle_{\rm atom} \right),
\end{equation}
where~$ |n\rangle $ is the Fock state of the cavity mode and
$ |{\tt g}\rangle \, ( |{\tt e}\rangle ) $ is
the ground (excited) state of the 2LA.

Here we are concerned with two-photon coincidence spectroscopy (2PCS)
which proceeds,
first by driving the atomic beam with a bichromatic field which
causes two-photon excitation to the second couplet of the JC ladder,
followed by two-photon decay from the atom-cavity system.
The objective is to count photon pairs emitted from the cavity
as the frequencies of the driving field are varied.
When the sum frequency is $2\omega\pm \sqrt2 g$, we expect to see
a resonantly enhanced two-photon count rate (2PCR). 
Of course, $g$ is a random variable due to beam fluctuations, and
this leads to inhomogeneous broadening.
Despite these difficulties, 2PCS appears to be a feasible
method for detecting the characteristic splitting in the JC 
model~\cite{Carmichael96,Sanders97}.
However, improvements in the procedure are important to ensure that 
the detection process is optimised.

In the following analysis we investigate the appropriate window time for 
counting photon pairs.
Photon pairs are not emitted from the cavity simultaneously due to the
randomness of photoemission from an optical cavity.
The detection of a photon pair thus depends on identifying a window
time~$\tau_{\rm w}$ such that,
for two photons detected with temporal separation~$t$ such that~$t < \tau_{\rm 
w}$,
the two photons are deemed to be members of a pair,
and, if~$t > \tau_{\rm w}$,
are deemed to be independent single photons (not members of a pair). 
Here we determine the optimal window time $\tau_{\rm opt}$ which maximises the 
counting rate of genuine pairs relative to the rate of false pair counts. 

\section{The Master Equation}
\label{sec:master}

The Hamiltonian (\ref{JCH}) for the combined atom-cavity system
ignores the driving field emission of photons from the system.
The Hamiltonian is useful in so far as it yields the spectrum for
the combined atom-cavity system, but the full quantum master
equation is necessary to calculate the quantities relevant
to experiments, namely the two-photon count rate (2PCR).
The experiment proceeds by measuring the 2PCR as a function of
the bichromatic driving field's scanning field frequency.

Two-photon excitation is provided by driving the atom directly 
with a bichromatic field, characterised by the time-dependent variable
\begin{equation} 
{\cal E}(t) = {\cal E}_1 e^{-i\omega_1 t} 
	+ {\cal E}_2 e^{-i\omega_2 t} .
\end{equation}
The angular frequency~$\omega_1$ is fixed and resonantly excites the
atom-cavity system from the ground state~$|0\rangle$ to the
excited state~$|1\rangle_-$  for the subensemble~$g=g_f$.
That is, provided that
\begin{equation}
\label{gf}
g_f = \omega - \omega_1 ,
\end{equation}
the bichromatic driving field will resonantly excite the 
subensemble of atom-cavity systems for which $g=g_f$.
Of course subensembles for which $g \neq g_f$ can also be
excited, but these excitations are non-resonant and hence
less significant.
The second frequency, $\omega_2$, is scanned over a particular frequency range.
The purpose of the second component of the
bichromatic field is to excite to one of the two states
in the second couplet of the Jaynes-Cummings ladder,
namely $|2\rangle_{\pm}$.
Thus, the range of scanning frequencies for $\omega_2$ must
include the $|1\rangle_- \longleftrightarrow |2\rangle_{\pm}$
transition frequencies,
\begin{equation}
\omega \pm ( \sqrt{2} \mp 1 ) g ,
\end{equation}
respectively.

The amplitudes of the two chromatic components must be large enough 
to ensure sufficient occupation of the excited state but not large 
enough that significant Stark shifting or nonnegligible occupation of the 
higher-order states occurs.
Enhanced rates of photon pair detection are then sought as the
scanning frequency~$\omega_2$ is varied.
The enhanced 2PCR occurs at
the resonances shown in Fig.\ \ref{fig:ladder}.

In addition to a driving term, 
loss terms must also be included.
The Born-Markov approximation is applied to both radiation reservoirs:
the reservoir for the field leaving the cavity and the reservoir
for direct fluorescence of the 2LA from the sides of the cavity.
The cavity damping rate is~$\kappa$,
and the emission rate into free space is~$\gamma$,
where~$\gamma$ is the inhibited spontaneous emission rate due to the 
restriction of modes by the cavity.
The master equation~\cite{Sanders97} can be expressed as
$\dot{\rho} = {\cal L}\rho$ for~$\cal L$ the Liouvillean superoperator.
More specifically the Liouvillean superoperator can be expressed 
as the sum of a time-independent Liouvillean operator,
a time-dependent Liouvillean operator and a `jump' term. By defining 
$\delta=\omega_2-\omega_1$, in the rotating 
picture of~${\cal E}_1$ the master equation is
\begin{equation}
\label{master}
\dot{\rho}(t;\delta,g)
	= \left[ {\cal L}(g) + {\cal L}(t;\delta) 
	+ {\cal J} \right] \rho(t;\delta,g)
\end{equation}
for
\begin{eqnarray}
\label{Heff}
H_{\rm eff}(g)
	&=& \left( \omega - \omega_1 \right) (\sigma_z + a^{\dagger} a) 
	+ i g (a^{\dagger} \sigma_- - a \sigma_+ )	\nonumber	\\
	&&+ i {\cal E}_1 ( \sigma_+ - \sigma_- )
	- i\kappa a^{\dagger} a - i(\gamma/2) \sigma_+ \sigma_-
\end{eqnarray}
a non-Hermitian Hamiltonian, the first term is

\begin{equation}
{\cal L}(g) \rho
	= -i \left[ H_{\rm eff}(g) \rho 
		- \rho H_{\rm eff}^{\dagger} (g) \right],
\end{equation}
the time-dependent term is 

\begin{equation}
\label{L(t)}
{\cal L}(t;\delta) \rho
	= {\cal E}_2 
		\left[ e^{-i\delta t} \sigma_+ 
		- e^{i\delta t} \sigma_- , \rho \right],
\end{equation}
and the jump term is
\begin{equation}
\label{jump}
{\cal J} \rho = \gamma \sigma_- \rho \sigma_+
	+ 2 \kappa a \rho a^{\dagger} .
\end{equation}

The atom-field coupling strength~$g$ depends on the atomic position~${\bf r}$.
Provided that the atoms move sufficiently slowly through the cavity 
\cite{Carmichael96,Sanders97},
the atom can be treated as if it were
at rest at some randomly located position~${\bf r}$.
The interaction of the atom with the cavity is then described by 
the master equation in the asymptotic large time ($t \longrightarrow \infty$). 
As the position~${\bf r}$ is a randomly varying quantity,   
the value of the coupling strength~$g$ itself is also random. 
Hence, a coupling strength distribution~$P(g)$ can be  
constructed~\cite{Sanders97}. 
The resultant density matrix is given by
\begin{equation}
\label{density:matrix}
\rho(t;\delta) = \int_{Fg_{\rm max}}^{g_{\rm max}} P(g) \rho(t;\delta,g) dg 
\end{equation}
where $g_{\rm max}$ is the coupling strength at a cavity node and 
$Fg_{\rm max}$ 
is the effective lower bound cut-off for the coupling $(0<F<1)$.
The effect of averaging over~$P(g)$ is an inhomogeneous spectral broadening.
This broadening is due to atomic position variability. In Fig.\ \ref{fig:P(g)}
two typical distributions~$P(g)$ are depicted, one for the case of 
a uniformly distributed atomic beam entering the cavity and the second
for an atomic beam initially passing through a rectangular 
mask~\cite{Sanders97}. In both cases we assumed a single-mode cavity
supporting a TEM$_{00}$ mode. 

For a bichromatic driving field, the density 
matrix~(\ref{density:matrix}) does not settle to a steady state value.
The time-dependence of the density matrix in the
long-time limit can be treated by making a Bloch function expansion
of the density matrix~\cite{Sanders97}.
In the Bloch function expansion,
the density matrix is written as the sum
\begin{equation}
\label{Bloch}
\rho(t) = \sum_{N=0}^{\infty} \rho_N(t) e^{\imath N \delta t}
\end{equation}
with~$\rho_N(t)$ time-dependent matrices.
In the long-time limit, $\dot{\rho}_N \approx 0$ and $\rho_N$ can 
thus be regarded as time-independent.
As the photocount integration time is expected to be long compared to the 
frequency~$\delta$, it is reasonable to approximate $\rho(t)$
by truncating the expansion~(\ref{Bloch}).

\section{The Two-Photon Count Rate} 
\label{sec:2PCR} 
 
The two-photon count rate (2PCR) can be obtained in more than one way.
Ideally one would have a perfectly efficient photodetector which
detects all photons leaving one side of the cavity.
The photodetector would then provide a complete record of photon
emissions from the cavity as a function of~$t$.
A perfect coincidence would then arise as two simultaneously 
detected photons at some time~$t$.
However, there are two problems.
One problem is that there does not exist a perfectly efficient photodetector.
Therefore, some pairs of photons are observed as single-photon emissions
because one member of the pair escapes observation.
In fact some pairs are missed altogether because both photons escape
detection.
The other problem concerns the detection of two simultaneously created photons.
Although created simultaneously, the emission from the cavity is
not simultaneous due to the randomness of the emission time
which is characterised by the cavity lifetime~$1/\kappa$.

A better and more accurate way to describe two-photon detections is to 
employ the 2PCR.
To begin with, we consider two photons to be coincident provided that they
arrive within a time interval~$\tau_{\rm w}$, the `window time'.
The choice of window time is not obvious, and it is our aim here
to determine what the window time should be.
As the two simultaneous photons can be separated by a time
of order $\kappa^{-1}$,
as discussed above,
the window time~$\tau_{\rm w}$ might be expected to be on the 
order of $\kappa^{-1}$.
However, our purpose here is to consider the choice of $\tau_{\rm w}$
in detail and to identify the optimal choice of window time~$\tau_{\rm w}$
which will produce the best measure of the 2PCR.

The choice of optimal window time is further complicated by the
method of detecting nearly simultaneous photons.
In the ideal case discussed above of a perfect photodetector
yielding a record of all photon emissions from the cavity,
one can then define a two-photon event as taking place if
a second photon arrives between times~$t_0$ and $t_0 + \tau_{\rm w}$,
{\em conditioned} on a photodetection at time~$t_0$.
We refer to this rate as the {\em conditional} 2PCR and
define this rate to be
\begin{eqnarray}
\label{2PCR:con}
\Delta^{(2)}_{\rm con}(\delta,g,\tau_{\rm w})
	&\equiv& 
	\lim_{t_0 \rightarrow \infty}\frac{1}{\tau_{\rm w}} \int_{t_0}^{t_0+\tau_{\rm w}} dt	\nonumber \\
	&& \times \left\langle : \hat{n}(t_0) \hat{n}(t_0+t) : 
	\right\rangle (\delta,g).
\end{eqnarray}
The number operator in eq~(\ref{2PCR:con}) is defined
as $\hat{n}(t) \equiv \hat{a}^{\dagger}(t) \hat{a}(t)$,
and `$: \, :$' refers to normal ordering.
The averaging is performed for the density matrix of eq~(\ref{master}).
The conditional 2PCR for a window time $\tau_{\rm w} = \kappa^{-1}$
was used in the quantum trajectory analysis of PCS in Ref.~\cite{Carmichael96}.

Another natural way to measure the 2PCR is by counting all photon pairs
defined as being separated by an interval less than~$\tau_{\rm w}$.
This 2PCR is referred to as the {\em unconditional} 2PCR and does
not rely on starting the count for the second photon conditioned
on detecting the first photon.
The definition of the unconditional 2PCR is
\begin{eqnarray} 
\label{2PCR:unc} 
\Delta^{(2)}_{\rm unc}(\delta,g,\tau_{\rm w})
	&=& \lim_{t_0 \rightarrow \infty}
	\frac{2}{\tau_{\rm w}^{2}} 
	\int_{t_0}^{t_0+\tau_{\rm w}}  dt^\prime 
	\int_{t_0}^{t^\prime}dt \nonumber \\
       && \times  \left\langle : \hat{n}(t)\
	\hat{n}(t^\prime) : \right\rangle (\delta,g). 
\end{eqnarray}
As shown in Appendix A, this expression can be simplified to read
\begin{eqnarray} 
\label{two:result} 
\Delta^{(2)}_{\rm unc}(\delta,g,\tau_{\rm w})
	= \frac{2}{\tau_{\rm w}^{2}}  
	\int_0^{\tau_{\rm w}} & du &  \int_0^{u} dw  \nonumber \\
	& \times &
	\left \langle : \hat{n}(0)\hat{n}(w) : \right \rangle (\delta,g).	
\end{eqnarray} 

We solve analytically for two extreme cases in Appendix A.
The window time can be extremely long
$(\kappa \tau_{\rm w} \gg 1)$,
yielding expression (\ref{long:time}),
or extremely short $(\kappa \tau_{\rm w} \ll 1)$,
yielding expression (\ref{short:time}) for both conditional and
unconditional 2PCR.
The short window time ($\tau_{\rm w} \longrightarrow$ 0) 
was the basis of the analysis of 2PCS in Ref.\ \cite{Sanders97}.
In this treatment both the conditional and unconditional 2PCR at time~$t$
is approximated by $\langle:\hat{n}^2(t):\rangle$. In the long-time limit
the 2PCR is dominated by Poissonian statistics. 

\section{The Optimal Window Time}
\label{sec:opt} 

The choice of optimal window time~$\tau_{\rm opt}$ depends on the
technique for observing two-photon coincidences, but another factor
must also be considered.
The purpose of 2PCS is to observe two-photon decay resonances
from the combined atom-cavity system.
As explained in Refs~\cite{Carmichael96,Sanders97},
there are three peaks in the 2PCR as a function of the scanning
frequency~$\delta$. These peaks are shown in Fig.~\ref{adadaa}
as a function of the normalised scanning field frequency 
\begin{equation}
\label{norm:delta}
\tilde\delta=\frac{\omega_2-\omega}{\omega-\omega_1}.
\end{equation}
The choice of $\tau_{\rm w}$ will depend on which peak is
being observed.
However, the best peak for observing a two-photon decay resonance
occurs for 
\begin{equation}
\label{delta:peak}
\tilde\delta = 1+\sqrt2.
\end{equation}
This resonance corresponds to the transition 
$|1)_- \! \longleftrightarrow~\!|2)_+$.
This peak does not occur in a semiclassical description of intensity
correlations in the cavity output field and therefore serves
as a signature of a genuine quantum field effect in CQED.
Moreover, this peak can be observed without the added complication of
having to perform the experiment twice, once with a bichromatic field
and once again with a monochromatic field, in order to perform the
signal enhancement technique of background subtraction~\cite{Carmichael96,Sanders97}.
Finally detection of this peak is the most feasible of the 
three dominant peaks in the two-photon spectrum.
Hence, we consider specifically $\tau_{\rm w}$ for this peak at
$\tilde\delta$ given by eq~(\ref{delta:peak}).

In Fig.~\ref{adadaa} we observe that the 2PCR peak sits on a background 2PCR
which is largely independent of~$\tilde\delta$ and occurs
at~$\overline{\langle : n^2 : \rangle} \approx 2.1\times 10^{-5}$.
Let us characterise the quality of this 2PCR peak by the ratio of the
peak height to the height of the background 2PCR.
We can understand this ratio in terms of signal to noise,
where signal is the 2PCR from the sought-for two-photon decay events,
and the background noise corresponds to two-photon decays arising
from unwanted off-resonance two-quantum excitations and decay events.
The peak-to-valley ratio (PVR) is determined by the height of the
peak to the height of the background (or valley) 2PCR.
The optimal window time
$\tau_{\rm w} = \tau_{\rm opt}$
is defined such that the PVR
for this 2PCR is maximal.
That is, either a larger or a smaller choice of the window time would
reduce the value of the PVR making the peak more difficult to detect.

There are other concerns besides the PVR in choosing the window time.
For example, choosing a much shorter window time could improve the
PVR but also lengthen the run time of the experiment in order to 
accumulate enough signal.
That is, the absolute height of the peak is also a matter of concern in 
determining the feasibility of the experiment and is determined by the
allowable timescale of the experiment. The minimum height would need to
be on the order of~$T^{-1}$ for~$T$ the timescale of the data collection.

The PVR is obtained numerically. The matrix continued fraction method is used
to solve the master equation to determine the peak height. The background, or
valley, can be solved analytically though by treating the detuning of
the scanning field as large. The details are provided in Appendix B. 

The 2PCR for large~$\tilde\delta$ is given
by expressions~(\ref{offres:c2PCS}) and~(\ref{offres:2PCS}).
The peak-to-valley ratio 2PCR is thus
\begin{equation}
\label{PVR}
{\rm PVR_{\xi}} = \frac{ \Delta^{(2)}_{\xi} (\tilde\delta, g, \tau_{\rm w}) }
	{ \left (\Delta^{(2)}_{\rm o} \right )_{\xi} (g, \tau_{\rm w} ) }
\end{equation}
where~$\xi\in\{\rm con, \rm unc\}$.
In Fig.\ \ref{fig:surfPVR} surface plots of the 
PVR {\em vs}~$g$ 
and~$\tau_{\rm w}$ reveals that the PVR increases as
$g$ decreases.
This is due to the background signal of two-photon 
coincidences for~$\tilde\delta$ large becoming negligible as shown in 
Fig.~\ref{adadaa}.
Although the PVR improves as~$g$ decreases,
the signal of two photon coincidences within the
window time~$\tau_{\rm w}$ decreases.
This decrease is due to the fact that the resonant
frequency for the transition
$|0)\longleftrightarrow |1)_-$ is~$\omega-g$, but we have
constrained the pump field frequency by~(\ref{gf}).
Hence, as~$g$ 
decreases, the pump field drives the system more and more off resonance.
The window time~$\tau_{\rm w}$ for achieving the optimal PVR is
an order of magnitude smaller than $\kappa^{-1}$. 
The optimal time~$\tau_{\rm opt}$ exhibited in Fig.~\ref{fig:optvsg} 
is a function of coupling strength and in Fig.~\ref{fig:resopt}
is a function of~$\gamma/\kappa$ (averaged over~$P(g)$).

We are concerned specifically with the 2PCR and the PVR for the system with
a coupling constant distribution based on the TEM$_{00}$ 
mode~\cite{Carmichael96,Sanders97}. The coupling strength
distribution is depicted in Fig.\ \ref{fig:P(g)}, and we treat the masked beam
case which enhances the large-coupling effect.

In Fig.~\ref{fig:avgPVR} 
we present the PVR for the density 
matrix~$\rho$ of expression~(\ref{density:matrix}), averaged 
over~$P(g)$. 
There is a peak
of the PVR for each $g/\kappa$ given by 
$\tau_{\rm w}=\tau_{\rm opt}$
(the optimal time window for observing the peak~(\ref{delta:peak})).
These values of~$\tau_{\rm opt}$ are plotted in Fig.\ \ref{fig:optvsg}
for $0\leq g/\kappa\leq 10$ for a range of values of~$\gamma$.
Of particular interest here is the very weak dependence of~$\tau_{\rm opt}$
on~$\gamma$ where~$\gamma$ is varied by a factor of~$20$. 
Moreover,~$\tau_{\rm opt}$ is generally decreasing as~$g/\kappa$ increases.

\section{Discussion}

In Fig.\ \ref{fig:avgPVR}~we observe a maximum of the 
PVR for each of the assumed~$P(g)$ in Fig.\ \ref{fig:P(g)}. This peak occurs
at~$\kappa\tau_{\rm opt}\approx 0.111$ for the conditional 2PCR and 
at~$\kappa\tau_{\rm opt}\approx 0.135$ for the unconditional 2PCR. 
We can understand the location of 
these peaks by referring to Fig.\ \ref{fig:optvsg}.

In Fig.\ \ref{fig:optvsg}(a)
the values of~$\kappa\tau_{\rm opt}$ for the conditional 2PCR
are predominantly between~$0.11$ 
and~$0.16$, but the value of~$\kappa\tau_{\rm opt}$ in the vicinity of $g/\kappa=9$
is between~$0.05$ and~$0.11$. Due to the resonance condition~(\ref{gf})
this region of the~$\kappa\tau_{\rm opt}$ {\em vs}~$g/\kappa$ curve is more
significant. 

Hence, the dependence of~$\tau_{\rm opt}$ on~$\gamma$ as depicted in
Fig.\ \ref{fig:resopt} is dominated by the~$g/\kappa=9$ region of 
Figs.\ \ref{fig:optvsg}.
Similarly, we can estimate the precise value of ~$\tau_{\rm opt}$ 
for the unconditional 2PCR from the dashed line of Fig.\ \ref{fig:resopt} 
in the
context of the values of~$\tau_{\rm opt}$ in 
Fig.\ \ref{fig:optvsg}(b).

The linear dependence of~$\tau_{\rm opt}$ on~$\gamma$, 
as seen in Fig.\ \ref{fig:resopt} yields
a correlation of 0.9983 for the conditional 2PCR, and
a correlation of 0.9995 for the unconditional 2PCR. 
In the conditional case

\begin{equation}
\label{av:cTopt}
\kappa \tau_{\rm opt}\approx-(1.4\times 10^{-3})\gamma/\kappa+
0.11 \, (\rm con),
\end{equation}
and in the unconditional case

\begin{equation}
\label{av:Topt}
\kappa \tau_{\rm opt}\approx-(2.1\times 10^{-3})\gamma/\kappa+
0.14 \,(\rm unc).
\end{equation} 
The low values of the slopes are indicative of the weak dependence on~$\gamma$.
Formulae (\ref{av:cTopt}) and (\ref{av:Topt}) can be used to fine-tune the
choice of optimal window time, and the linear relationship 
simplifies the task of interpolating to 
obtain~$\tau_{\rm opt}$. 
 
\section{Conclusions} 
\label{sec:conclusions}
We have determined expressions for the optimal window time~$\tau_{\rm opt}$ 
for both conditional and unconditional 2PCR.
These expression provide an optimal PVR
for the 2PCR peak at~$\tilde\delta=1+\sqrt{2}$
corresponding to the~$|1)_-\longrightarrow |2)_+$ transition. Although
we have determined~$\tau_{\rm opt}$ for certain parameters and for the
coupling-strength distribution~$P(g)$ (solid line of Fig.\ \ref{fig:P(g)}), the
algorithm presented here is sufficiently general to allow calculation 
of~$\tau_{\rm opt}$ for other parameters and other coupling-strength
distributions. In general~$\tau_{\rm opt}$ is smaller than~$\kappa^{-1}$
by an order of magnitude for both the conditional and unconditional 2PCR. 
There is some dependence on~$\gamma$, but this
dependence is weak and is close to linear in the cases studied here. 

Analyses of optimal window times are aided by studies of~$\tau_{\rm opt}$ 
for particular values of~$g$, that is, for the coupling-strength
distribution~$P(g)$ corresponding to~$\delta(g-g_0)$ for some~$g_0$. 
These calculations provide good estimates of the optimal window time
for general~$P(g)$.
A longer window time may be desirable, however, if the timescale
for collecting enough data is not experimentally feasible.
A compromise between the two objectives of optimising the PVR and of 
accumulating sufficient data to produce a large peak height may be necessary.

An important technique discussed in Refs.\ \cite{Carmichael96,Sanders97}
was background subtraction. The principle behind this method is to remove
the unwanted two-photon off-resonance excitation to the second couplet.
Background subtraction is particularly important to improve the PVR for 
2PCR peaks.
However, we choose to study detection of the most promising
peak experimentally~(\ref{delta:peak}).
In our simulations we determine the PVR for this peak both by performing
background subtraction and without background subtraction, and we obtain  
plots in agreement with
Figs.\ \ref{fig:surfPVR} and~\ref{fig:avgPVR}.
These results confirm
the assertion in Ref.\ \cite{Sanders97} that ``the resonance frequency lies
outside the inhomogeneous line and the resonance should be resolved even in
the presence of the two-photon background''. Hence, background subtraction
is not necessary to obtain optimal window times for resolving the
peak~(\ref{delta:peak}).

\section*{Acknowledgements}
Dr S.\ M.\ Tan has provided Matlab$^{\rm TM}$ programs to us which 
was used initially 
to double check our simulations of the monochromatically driven
Jaynes-Cummings system. We have benefited from valuable
discussions with H.\ J.\ Carmichael, J.\ D.\ Cresser, Z.\ Ficek,
K.-P.\ Marzlin, and S.\ M.\ Tan. This research has been supported by
Australian Research Council Large and Small Grants and a Macquarie
University Research Grant.
\newpage 

\appendix
\section*{A: The conditional and unconditional two-photon count rate (2PCR)}
\label{app:rate} 
 
In the long-time limit,
the conditional two-photon count rate (2PCR) is given by
\begin{eqnarray}
\label{two:photo:b}
\Delta^{(2)}_{\rm con} (\delta, g, \tau_{\rm w}) 
= \frac{1}{\tau_{\rm w}}\int_{0}^{\tau_{\rm w}} dt
\left \langle : \hat{n}(0) \hat{n}(t): \right \rangle(\delta, g).
\end{eqnarray}
If the time window $\tau_{\rm w}$ is large,
compared to $\kappa^{-1}$ (the cavity lifetime), the two photons are highly 
decorrelated, and we can approximate
\begin{equation} 
	\label{long:time:exp} 
	\left\langle : \hat{n}(0) \hat{n}(t) : \right\rangle(\delta,g)
	\longrightarrow \left\langle \hat{n}(0) \right\rangle^{2}(\delta,g). 
\end{equation} 
Thus,
\begin{equation} 
\label{long:time} 
\Delta^{(2)}_{\rm con}(\delta,g, \tau_{\rm w}) \longrightarrow 
	\left\langle \hat{n}(0) \right\rangle^{2}(\delta,g). 
\end{equation}
This count rate reflects the Poissonian nature of the count statistics for 
long window times. 
On the other hand, for $ \kappa \tau_{\rm w} \ll 1$, 
the correlation between photon pairs cannot be neglected. Hence, 
the count rate reduces to 
\begin{eqnarray}	 
\label{short:time} 
\Delta^{(2)}_{\rm con}(\delta,g, \tau_{\rm w})\longrightarrow 
	\left \langle : \hat{n}^2(0) : \right\rangle (\delta,g)
\end{eqnarray} 
which is the approximation employed in Ref~\cite{Sanders97}.

Similarly, in the long-time limit, the unconditional 2PCR 
is

\begin{equation} 
\label{two:photo:a} 
\Delta^{(2)}_{\rm unc}\left( \delta,g,\tau_{\rm w} \right)
	= \frac{2}{\tau_{\rm w}^{2}} \int_0^{\tau_{\rm w}} dt^\prime 
	\int_0^{t^\prime}dt 
	\left\langle : \hat{n}(t)\hat{n}(t^\prime) : \right \rangle
	\left( \delta, g \right).
\end{equation} 
This expression can be simplified as we show below.
First we make the substitution~$u_{\pm} = (t^\prime \pm t)/\sqrt{2}$.
We also introduce the notation~$d^2 u=du_- du_+$ and let~$\nu$ be the union
of the two regions~$\{0 < u_- < \tau_{\rm w}/\sqrt2, 0 < u_+ < u_-\}$ 
and~$\{\tau_{\rm w}/\sqrt2 < u_- < \sqrt2 \tau_{\rm w},
0 < u_+ < \sqrt2\tau_w-u_-\}$

This substitution transforms the above double integral into 
the sum of two double integrals:
\begin{eqnarray} 
\label{two:arrange} 
	\Delta^{(2)}_{\rm unc}(\delta,g,\tau_{\rm w}) &=&  
	\frac{2}{\tau_{\rm w}^{2}}  
	\int\int_{\nu}
	\left\langle : \hat{n} \left(\frac{u_{+}-u_{-}}{\sqrt{2}}\right) 
	\hat{n} \left(\frac{u_{+}+u_{-}}{\sqrt{2}}\right) :
	\right\rangle (\delta,g) \nonumber \\
	& = & \frac{2}{\tau_{\rm w}^{2}}  
	\int\int_{\nu}
	\left \langle : \hat{n}(0) \hat{n}(\sqrt{2}u_{+}) : \right\rangle
	(\delta,g).  
\end{eqnarray}
The advantage of this expression is that the two-time photon
number correlation depends on only one term in the double 
integral instead of both terms in the double integral.

Greater simplification is possible and desirable for studying the
short and long window time~$\tau_{\rm w}$.
Substituting $u_{\pm}= w_{\pm}/\sqrt{2}$ transforms 
eq~(\ref{two:arrange}) to
\begin{eqnarray} 
\label{two:arrange:one} 
\Delta^{(2)}_{\rm unc}(\delta,g,\tau_{\rm w}) & = & 
	\frac{1}{\tau_{\rm w}^{2}} \Bigg[ 
	\int_0^{\tau_{\rm w}} dw_- 
	\int_0^{w_-} dw_+
		\nonumber	\\ &&+
	\int_{\tau_{\rm w}}^{2\tau_{\rm w}} dw_- 
	\int_0^{2\tau_{\rm w}-w_-} dw_+ \Bigg] 
		\nonumber \\ &&
	\left\langle : \hat{n}(0) \hat{n} (w_+) : \right\rangle (\delta,g)
\end{eqnarray} 
which reduces to 
\begin{eqnarray} 
\label{two:res} 
\Delta^{(2)}_{\rm unc}(\delta,g,\tau_{\rm w}) & = &
	\frac{2}{\tau_{\rm w}^{2}} \int_0^{\tau_{\rm w}} du 
        \int_0^{u} dw \nonumber \\ 
	& & \times \left\langle : \hat{n}(0) \hat{n} (w) : \right\rangle (\delta,g) .
\end{eqnarray}

For large ($\tau_{\rm w} \gg \kappa^{-1}$) and small 
($\tau_{\rm w} \ll \kappa^{-1}$) window times~$\Delta^{(2)}_{\rm unc}$ reduces 
identically to~$\Delta^{(2)}_{\rm con}$ as shown 
in equations~(\ref{long:time}) and (\ref{short:time}).

\section*{B: Background of conditional and unconditional 2PCR}

For the scanning field far off resonance ($\delta$ large),
the time-dependent component of the Liouvillean~(\ref{L(t)})
can be ignored.
In the interaction picture, 
the master equation can be written as~$ \dot{\rho} = {\cal L} \rho $
with $\cal L$ time-independent.
If $\rho$ is expressed as a vector, then~$\cal L$ can be expressed 
as a complex matrix with~$ \{ - \lambda_n | n \in {\cal Z}_{N^2} \} $
the set of eigenvalues for~$N$
the number of levels in the Jaynes-Cummings ladder retained after truncation.
The density matrix can be approximated by the sum
\begin{equation}
\label{rhosum}
\rho(t) = \sum_{n=1}^{N^2}  \rho_n e^{- \lambda_n (t-t_0)}
\end{equation}
for~$ \{ \rho_n \} $ a set of time-independent $N \times N$ matrices. Thus, 
the conditional 2PCR~(\ref{2PCR:con}) can be written as 
\begin{eqnarray} 
\label{offres:c2PCS} 
\left (\Delta^{(2)}_{\rm o}\right )_{\rm con}(g, \tau_{\rm w}) 
	& = &  c_0 (g)+ \frac{1}{\tau_{\rm w}}  
	\int_0^{\tau_{\rm w}} dt \nonumber \\	
	& \times &  \sum_{n=1}^{N^2} c_n(g) 
	\exp{\left[ -\lambda_n (g) t \right]} 
	\nonumber \\
	& = & \! c_0(g) \!
	\!  + \! \sum_{n=1}^{N^2} \! \frac{ c_n (g) } { \mu_n(g) }\!
	\left \{ \! 1- e^{ -\mu_n (g) \!} \right \} \! \! .
\end{eqnarray}
with $N^2$ scalar constants~$\left \{ c_n(g) \right \}$
and $N^2$ constants~$\left \{ \mu_n(g) \right \}$ where
\begin{equation}
\mu_n (g) = \lambda_n (g) \tau_{\rm w}  \neq 0, \hspace{0.2cm}
{\rm Re}\left \{\lambda_n(g)\right \} \geq 0
\end{equation}
as well. In the long window time limit,
we equate
\begin{equation}
\label{limit:Delta}
c_0 (g) = \lim_{\tau_{\rm w} \longrightarrow \infty} 
\left(\Delta^{(2)}_{\rm o} \right )_{\rm con}
(g,\tau_{\rm w}).
\end{equation}
Expansion~(\ref{offres:c2PCS}) provides a useful method for calculating 
$\left (\Delta^{(2)}_{\rm o} \right)_{\rm con}(g,\tau_{\rm w})$. 
The function $\left ( \Delta^{(2)}_{\rm o} \right)_{\rm con}(g,\tau_{\rm w})$ 
is monotonically increasing because 
$\partial \left (\Delta^{(2)}_{\rm o} \right )_{\rm con}
(g,\tau_{\rm w})/\partial \tau_{w} > 0$ if $\tau_{\rm w} \longrightarrow 
\infty$. 
Thus, $\partial \left(\Delta^{(2)}_{\rm o}\right)_{\rm con}
(g,\tau_{\rm w})/\partial \tau_{w}
\longrightarrow 0$ as the function approaches the limit given by 
(\ref{limit:Delta}).

In the same way, the unconditional 2PCR can be obtained:
\begin{eqnarray} 
\label{offres:2PCS} 
\left(\Delta^{(2)}_{\rm o} \right)_{\rm unc}(g, \tau_{\rm w}) 
	& = &  c_0 (g)+ \frac{2}{\tau_{\rm w}^{2}}  
	\int_0^{\tau_{\rm w}}  
	du \int_0^{u} dw  \nonumber \\
	& &\times \sum_{n=1}^{N^2} c_n(g) \exp{\left[ -\lambda_n (g) w \right]} 
	=c_0(g)	\nonumber \\ 
	  && + \! 2 \! \sum_{n=1}^{N^2} \frac{ c_n (g) } { \mu_n (g) }\!
	\left\{ \! \frac{ e^{ -\mu_n (g) } - 1}{ \mu_n (g) }+ 1 \! \right\}\! . 
\end{eqnarray} 

Thus, as in~(\ref{limit:Delta}), the long time limit reduces~(\ref{offres:2PCS}) to~$c_0(g)$.

\newpage

\begin{figure} 
	\caption{ 
	Selection of two subpopulations of the inhomogeneously broadened
	Jaynes-Cummings (JC) system via distinct absorption paths.
	The ground state energy and the inhomogeneously broadened energy bands
	associated with the first, second and third couplet of the 
	JC spectrum are shown.
	}
\label{fig:ladder} 
\end{figure}

\begin{figure}
	\caption{The scaled coupling strength distributions~$\kappa P(g)$ 
	as a function of~$g/\kappa$ for single atoms passing
	through an optical cavity supporting a single TEM$_{00}$ mode.
	The solid curve corresponds to a typical distribution for a
	rectangular mask filtering the atomic beam. The dashed line
	corresponds to the absence of a mask.}
\label{fig:P(g)}
\end{figure}

\begin{figure}
	\caption{
	Spectrum of the simultaneous 2PCR (averaged over the 
	coupling strength distribution
	$P(g)$ shown with the solid line in Fig.\ \ref{fig:P(g)}) 
	as a function of the normalised driving field 
	frequency~$\tilde\delta$.
	}
\label{adadaa}
\end{figure}

\begin{figure}
	\caption{
	The peak-to-valley ratio 
	of the (a) conditional (PVR$_{\rm con}$) and 
	(b) unconditional (PVR$_{\rm unc})$ 2PCR over the scaled
	coupling strength~$g/\kappa$ and the scaled window 
	time~$\kappa\tau_{\rm w}$
	for the scaled loss rate~$\gamma/\kappa=2$.}
	\label{fig:surfPVR}
\end{figure}

\begin{figure}
	\caption{ 
	The scaled optimal window time $\kappa \tau_{\rm opt}$ 
	{\em vs} the scaled coupling strength $g/\kappa$ for
	$\diamond$ $\gamma/\kappa=0.2$, $+$ $\gamma/\kappa=2$,
	$\Box$ $\gamma/\kappa = 5$,
	$\times$ $\gamma/\kappa=7$ and
	$\triangle$ $\gamma/\kappa=10$ for (a) the conditional 
	and (b) the unconditional 2PCR.}
	\label{fig:optvsg}
\end{figure}

\begin{figure}
	\caption{
	The scaled optimal window time 
	$\kappa\tau_{\rm opt}$ {\em vs} the scaled loss rate~$\gamma/\kappa$  
	for the masked atomic beam. The symbol $+$ corresponds to the
	conditional~$\tau_{\rm opt}$ and~$\diamond$ to the 
	unconditional~$\tau_{\rm opt}$. Linear regression methods yield the
	two lines. For the conditional case, the slope is~$-1.4\times10^{-3}$, 
	the intercept is~$0.11$, and the correlation coefficient 
	is~$r=-0.9983$. 
	For the unconditional case, the slope is~$-2.1\times10^{-3}$, 
	the intercept is~$0.14$, and the correlation coefficient 
	is~$r=-0.9995$.   
	}
\label{fig:resopt} 
\end{figure}

\begin{figure}
	\caption{
	The  peak-to-valley ratio (PVR) of the conditional (solid line) 
	and unconditional (dashed line) 2PCR 
	(for the masked atomic beam) {\em vs} 
	the scaled window time~$\kappa\tau_{\rm w}$
	for the scaled loss rate $\gamma/\kappa=2$.
	}  
\label{fig:avgPVR} 
\end{figure}

\end{document}